\documentclass[reprint,amsmath,amssymb,braket,aps,prb]{revtex4-2}
\usepackage{graphicx,color,braket}% Include figure files
\usepackage{dcolumn}% Align table columns on decimal point
\usepackage{bm}% bold math
\usepackage[colorlinks=true,linkcolor=blue]{hyperref}
\usepackage{colortbl}
\usepackage[russian]{babel}
\newcommand{\rev}[1]{\textcolor{black}{#1}}
\newcommand{\rrev}[1]{\textcolor{black}{#1}}
\begin{document}
	
	\title{Exact theory of edge diffraction and launching of transverse electric plasmons at two-dimensional junctions
%	Точная теория краевой дифракции и возбуждения поперечных электрических плазмонов на двумерных контактах
	}
	
	\author{Dmitry Svintsov}
	\affiliation{Laboratory of 2d Materials for Optoelectronics, Moscow Institute of Physics and Technology, Dolgoprudny 141700, Russia}
	\email{svintcov.da@mipt.ru}
	
	\author{Alexander Shabanov}
	\affiliation{Laboratory of 2d Materials for Optoelectronics, Moscow Institute of Physics and Technology, Dolgoprudny 141700, Russia}

	\begin{abstract}
		An exact solution for electromagnetic wave diffraction at the junction of two-dimensional electron systems (2DES) is obtained and analyzed for electric field polarized along the edge. A special emphasis is paid to the metal-contacted and terminated edges. In the former case, electric field at the edge tends to zero; in the latter case, it tends to a finite value which is screened by 2d system in an anomalous fashion. For both types of edge and capacitive type of 2d conductivity, an incident wave excites transverse electric 2d plasmons. The amplitude of excited TE plasmons is maximized and becomes order of incident wave amplitude for capacitive impedance of 2DES order of free space impedance. For both large and small 2DES impedance, the amplitude of TE plasmons tends to zero according to the power laws which are explicitly derived.
	\end{abstract}
	
	\maketitle
	
	Junctions of two-dimensional electron systems (2DES) their contacts with metals are among the central objects in the 2D optoelectronics. Such junctions are capable to generate photocurrent~\cite{Tielrooij_Photocurrent_NNano,Muravev_spectrometer,Titova_PNJ_BLG,Olbrich_Ratchet}, which ensures their rich applications both in practical optoelectronics and fundamental studies of light-matter interactions. While numerous studies were devoted to the microscopic theories of electronic transport at the junctions in external electromagnetic (EM) fields~\cite{Echtermeyer_MGJ,Levitov_Hot_Theory,Nalitov_Ratchet,Ratchet_Hydro}, much less is known about diffraction of the EM fields at these junctions. The problem of EM wave diffraction at the laterally non-uniform 2DES is complex and is typically studied with electromagnetic simulations~\cite{Fateev_Rectification,Fateev_Transformation} or approximate plane wave representations of the local fields~\cite{Aizin_Finite_PCs,Gorbenko_Lateral_PCs}. Quite recently, it was realized that a powerful analytical method for solving integral scattering-type equations in the semi-infinite domains, the Wiener-Hopf method, can be successfully applied to the edge diffraction in 2DES~\cite{Margetis_WH,China_WH,Scattering_by_conductivity_contrast_exact,Nikulin_Edge,Alymov_Fresnel}. Several remarkable analytical results have been obtained using this technique. These include a universal value of electromagnetic absorbance at the metal-2DES contact~\cite{Nikulin_Edge}, a universal value of 2D plasmons' amplitude launched at the terminated 2D edge~\cite{China_WH}, but are not limited to the latter. 
	
	All previous studies of the edge diffraction in 2DES~\cite{Margetis_WH,Nikulin_Edge,China_WH} dealt with incident fields having the magnetic vector ${\bf H}_0$ along the edge of 2DES. Consequently, the electric field ${\bf E}_0$ was orthogonal to edge.  For such a polarization, the local electric field ${\bf E}(x)$ was greatly enhanced via the dynamic lightning-rod effect. Such enhancement has been confirmed experimentally via studies of polarization-dependent photocurrent at metal-graphene junctions~\cite{Tielrooij_Photocurrent,Semkin_APL,Semkin_NL}. 
	
	Another polarization of the incident field, where the electric vector ${\bf E}_0$ is directed along the edge, did not yet gain attention in the theory of 2D edge diffraction. The present paper fills in this gap. For such polarization, electric field is expected to be suppressed, especially if a 2DES is contacting a metal with very large conductivity. While field suppression is not as appealing as field enhancement, its particular magnitude is important for design of polarization-sensitive and polarization-resolving photodetectors~\cite{Semkin_APL,Wei_config_polarity}. Another intriguing aspect of such polarization is the possibility to excite the two-dimensional transverse electric (TE) plasmons~\cite{Mikhailov_new_mode} in such a simple scattering geometry. Such waves can exist only for capacitive type of 2DES conductivity $\sigma = \sigma'+i\sigma''$, $\sigma'' < 0$ if the notation $e^{-i\omega t}$ for time-dependent fields is assumed. \rrev{These waves do not accumulate electric charge upon propagation, therefore being different from conventional longitudinal plasmons existing for $\sigma''>0$.} While the conductivity of free-carrier interband motion is always inductive, the capacitive conductivity can be expected near the interband absorption edges~\cite{Kotov_Universal_TE}. Particularly, this can occur for photon energies $\hbar\omega$ close to the band gap $E_g$ (if present)~\cite{Kotov_Universal_TE} or close to the twice the Fermi energy, $2\mu$, in the case of doped graphene~\cite{Mikhailov_new_mode,TE_Experment}. \rrev{A formal solution of the diffraction problem in both polarizations for terminated imperfectly conducting sheet can be found in older Refs.~\cite{Senior,senior_oblique} . However, the launching of TE plasmons for inductive conductivity was not realized in these papers, presumably due to irrelevance of this case to classical metals. The case of diffraction at a junction of dissimilar materials was also not addressed.}
	
	We proceed to obtain an analytical solution for EM wave diffraction at the junction between 2D electron systems in TE polarization, i.e. for electric vector of the wave directed along the edge. Our particular emphasis will be on the asymptotic values of the field near the junction and away from it, as well on the amplitude of edge-launched TE plasmons.
	
\begin{figure}[ht!]
		\centering
		\includegraphics[width=0.9\linewidth]{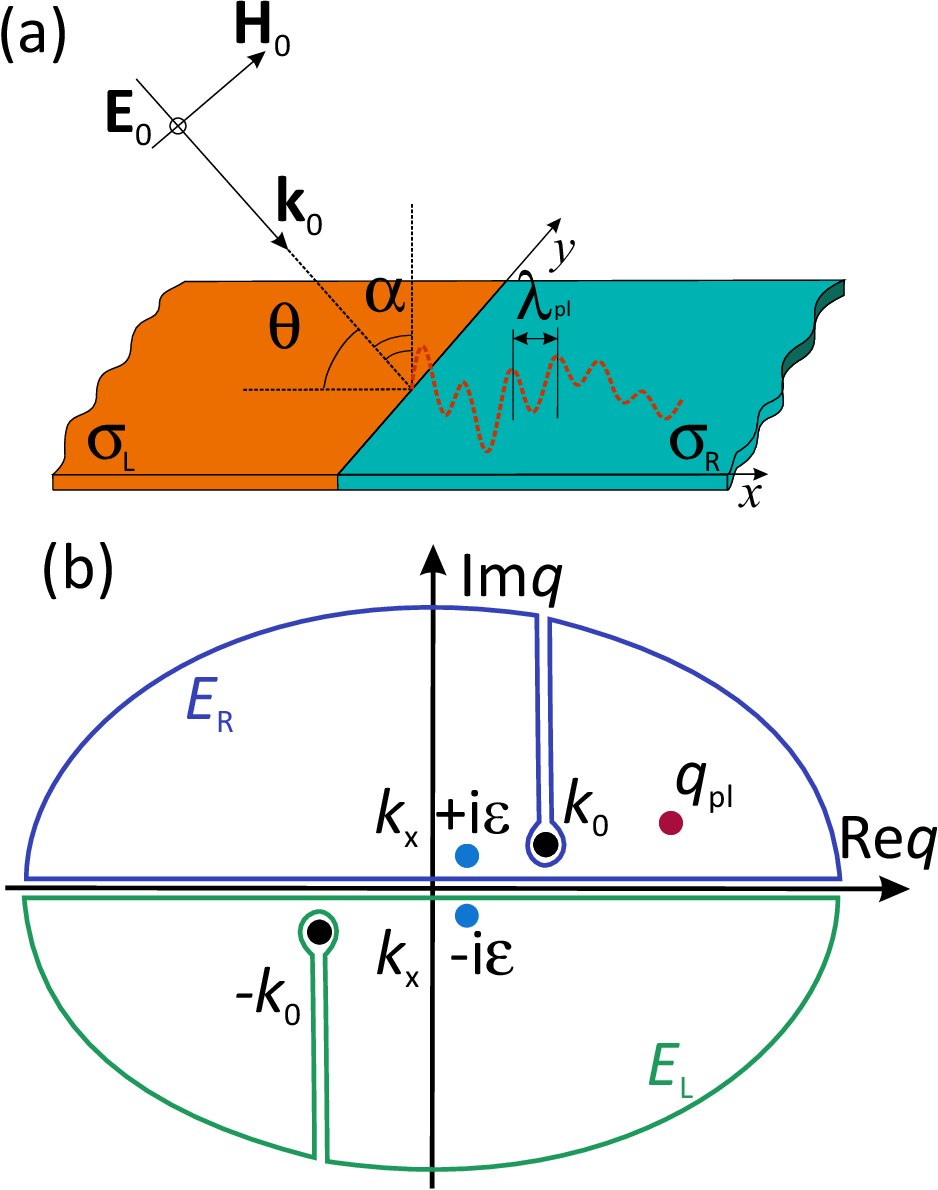}
		\caption{(a) Setup of the scattering problem: an $s$-polarized electromagnetic wave with wave vector $k_0$ is incident on the 2DES junction with conductivities $\sigma_L$ and $\sigma_R$ at gliding angle $\theta$. Schematic view of coordinate-dependent diffracted electric field is shown with red dashed line; the field can possess TE-plasmonic component with wavelength $\lambda_{pl}$ (b) Analytical structure of the scattering equation in the complex $q$-plane. Poles of the incident field appear at $q = k_x \pm i \epsilon$, branch points of the dielectric functions $\varepsilon(q)$ start at $q=k_0$ and run to infinity. Extra plasma poles can be present, depending on the sign of 2d conductivities $\eta''_{L/R}$}
		\label{fig:structure}
	\end{figure}

	The setting under study is shown in Fig.~\ref{fig:structure} (a). 2D electron systems with conductivities $\sigma_L$ and $\sigma_R$ lie in the plane $z=0$ and are contacting along a straight line $x=0$. Electric field of the incident wave is directed along the $y$-axis and has the form $E_y (x)= E_0 e^{i k_x x}$, here $k_x = k_0 \cos\theta$ is the $x$-component of the wave vector, $k_0$ is the wave number, and $\theta$ is the gliding angle. The harmonic time dependence of all quantities $e^{-i\omega t}$ will be assumed and hitherto skipped, so  will be skipped the subscript for $y$-component. The scattering equation for electric field in the 2DES plane $z=0$ has the form (see Supplementary material, section I, for derivation)
	\begin{equation}
		\label{eq-scattering}
		{E}\left( x \right)={{E}_{0}}{{e}^{i{{k}_{x}}x}}-\frac{Z_0 k_0}{4}\int\limits_{-\infty }^{+\infty }{{{H}_{0}}\left( {{k}_{0}}\left| x-{x}' \right| \right){j}\left( {{x}'} \right)d{x}'},
	\end{equation}
	where $Z_0$ is the free-space impedance ($4\pi/c$ in Gaussian units or $\sqrt{\mu_0/\varepsilon_0}\approx 377$ Ohm in SI units); ${H}_{0}$, the Hankel function of the zeroth order, is the fundamental solution of the wave equation in two dimensions, and ${j}\left( x \right)$ is the distribution of wave-induced surface currents in 2DES. To form a complete scattering equation with one unknown function, the electric field $E(x)$, we use the Ohm's law in the local form ${j}\left( x \right) = \sigma(x) E(x)$, where the conductivity varies in a stepwise fashion $\sigma\left( x \right)={{\sigma }_{L}}\theta \left( -x \right)+{{\sigma }_{R}}\theta \left( x \right)$.
	
	Solution of (\ref{eq-scattering}) relies on the Wiener-Hopf technique. It involves the splitting of the total field into the left and right components, $E\left( x \right)=E_{L}(x) \theta \left( -x \right)+E_{R}(x) \theta \left( x \right)$, Fourier transform of the governing equation with passing to the wave vector $q$-variable, and studies of the emerging functions of complex $q$-variable. \rev{The Fourier transformed fields $E_{L}(q)$ and $E_{R}(q)$ will be analytic in the upper and lower  half-planes of the complex $q$-variable, respectively.} Before actual Fourier transforming, it is reasonable to 'bound' all functions to a finite region of the real space. We replace the source field according to ${{E}_{0}}{{e}^{i{{k}_{x}}x}}\to {{E}_{0}}{{e}^{i{{k}_{x}}x}}{{e}^{-\epsilon \left| x \right|}}$, which mimics constraining the incident beam by the aperture of size $\sim \epsilon^{-1}$. We also assume some residual dissipation in the materials surrounding the 2DES, which amounts to the replacement $k_0 \to k_0 (1+i\delta)$, $\delta \ll 1$, in the argument of the electromagnetic propagator. We shall see that the final results will not depend on auxiliary values of $\epsilon$ and $\delta$. Still, retaining their finite magnitude is important in the course of solution.
	
	The Fourier transformed scattering equation (\ref{eq-scattering}) has the form
	\begin{multline}
		\label{eq-scattering-fourier}
		\varepsilon _{L} \left( q \right){{E}_{L}}\left( q \right)+\varepsilon _{R} \left( q \right){{E}_{R}}\left( q \right)=
		\\{{E}_{0}}\left[ 
		\frac{-i}{q-\left( {{k}_{x}}+i\epsilon  \right)}
		+
		\frac{i}{q-\left( {{k}_{x}}-i\epsilon  \right)} \right],
	\end{multline}
	where the transverse-electric screening functions are
	\begin{equation}
		\label{eq-dielectric-function}
		\varepsilon_{i} \left( q \right)=1+\eta_i\frac{{{k}_{0}}}{\sqrt{k_{0}^{2}-{{q}^{2}}}}, \quad \eta_i = \frac{Z_0 {{\sigma }_{i}}}{2}, \quad i = \{L,R\}.
	\end{equation}
	\rev{The branch cuts of the dielectric functions $\varepsilon_i(q)$ start at $q=\pm k_0$ and run to $\pm i\infty$, respectively, without crossing the real axis. Such choice is dictated by the decaying character of the electromagnetic propagator $H_0(k_0|x-x'|)$ at large distances.} The key role in the subsequent solution will be played by the 'factorized functions' $\varepsilon _{i+} \left( q \right)$ and $\varepsilon _{i-} \left( q \right)$. They are analytic in the upper and lower half-planes of the complex $q$-variable. The factorization here will be achieved with the Cauchy theorem applied to a narrow strip enclosing the real axis, though semi-analytical approaches are also available~\cite{Nikulin_Edge,Margetis_analytical}. The result of Cauchy factorization reads as
	\begin{equation}
		\label{eq-cauchy-factorization}
		\varepsilon _{i,\pm } \left( q \right)=\exp \left\{ \pm \frac{1}{2\pi i}\int\limits_{-\infty }^{+\infty }{\frac{\ln \varepsilon _{i} \left( u \right)du}{u-\left( q\pm i\gamma  \right)}} \right\}.
	\end{equation}
	After algebraic rearrangements, described in detail in Supplementary section II, Eq.~(\ref{eq-scattering-fourier}) takes the form
	\begin{multline}
		\label{eq-scattering-factorized}
		\frac{\varepsilon _{L+} \left( q \right)}{\varepsilon _{R+} \left( q \right)}{{E}_{L}}\left( q \right)+\frac{\varepsilon _{R-} \left( q \right)}{\varepsilon _{L-} \left( q \right)}{{E}_{R}}\left( q \right)=\\
		\frac{{{E}_{0}}}{\varepsilon _{R+} \left( {{k}_{x}} \right)\varepsilon _{L-} \left( {{k}_{x}} \right)}
		\left[ 
		\frac{i}{q-\left( {{k}_{x}}-i\epsilon  \right)} +
		\frac{-i}{q-\left( {{k}_{x}}+i\epsilon  \right)}
		\right],
	\end{multline}
	where the first terms of lhs and rhs are analytic in the upper half-plane, and the second terms of lhs and rhs are analytic in the lower half-plane. Therefore, one can equate the function analytic in the respective half-planes term-by-term. This results in the final solution for electric fields
	\begin{gather}
		\label{eq-solution}
		{{E}_{L}}\left( q \right)=\frac{+i{{E}_{0}}}{1+{{\eta }_{L}}/\sin \theta }\frac{\varepsilon _{R+} \left( q \right)}{\varepsilon _{L+}\left( q \right)}\frac{\varepsilon _{L+}\left( {{k}_{x}} \right)}{\varepsilon _{R+}\left( {{k}_{x}} \right)}\frac{1}{q-\left( {{k}_{x}}-i\epsilon  \right)}, \\ 
		{{E}_{R}}\left( q \right)=\frac{-i{{E}_{0}}}{1+{{\eta }_{R}}/\sin \theta }\frac{\varepsilon _{L-} \left( q \right)}{\varepsilon _{R-} \left( q \right)}\frac{\varepsilon _{R-} \left( {{k}_{x}} \right)}{\varepsilon _{L-} \left( {{k}_{x}} \right)}\frac{1}{q-\left( {{k}_{x}}+i\epsilon  \right)} .
	\end{gather} 
	Equations (\ref{eq-solution}) solve the scattering problem in the Fourier domain. \rev{They agree well with results of electromagnetic simulations performed using the CST Microwave studio package, as demonstrated in Supplementary section III.} \rrev{The solution applies to the 2DES plane $z=0$. The solution outsude this plane can be obtained by multiplying by the 'phase exponent' $E(q,z) = [E_L(q,z=0) + E_R(q,z=0)]e^{i\sqrt{k_0^2-q^2}|z|}$.}
	
	\rrev{We shall limit our discussion to the terminated ($\eta_{L}=0$) and metal-contacted ($|\eta_{L}| \gg  1$) edges. The precise condition of 'good metallicity' $|\eta_{L}| \gg  1$ depends on the metal thickness $t$. Once $t$ is above the skin depth, $\eta_{L}$ is the product of bulk conductivity and skin depth, which is nothing but metal refractive index $n_M$. Except for ultraviolet range, $|n_M| \gg 1$. Once $t$ is below the skin depth, $\eta_{L}$ is the product of bulk metal conductivity in units of $c/2\pi$ and its thickness $t$. It is possible to show that even for $t\sim 10$ nm and $\omega/2\pi \sim 1$ THz we have $|\eta_{L}| \sim  10^2$ for metals like gold and copper.} The fields for terminated 2DES and metal-contacted 2DES will be equipped with symbols $\times$ and $\leftrightarrow$, respectively. The subscript $R$ of the 'right' material functions will be skipped, $\varepsilon_{R} \equiv \varepsilon$, $\eta_{R} \equiv \eta$. After taking the proper limit of the 'left' dielectric function, we arrive at the explicit form of 'right' electric field for both types of the edge:
	\begin{gather}
		\label{eq-er-terminated}
		E_{R}^{\times}\left( q \right)=\frac{-i{{E}_{0}}}{1+{\eta}/\sin \theta }\frac{\varepsilon _{-}\left( {{k}_{x}} \right)}{\varepsilon _{-}\left( q \right)}\frac{1}{q-\left( {{k}_{x}}+i\epsilon  \right)},\\
		\label{eq-er-metallized}
		E_{R}^{\leftrightarrow}\left( q \right)=\frac{-i{{E}_{0}}}{1+{{\eta }}/\sin \theta }\frac{\sqrt{{{k}_{0}}-{{k}_{x}}}}{\sqrt{{{k}_{0}}-q}}\frac{\varepsilon _{-}\left( {{k}_{x}} \right)}{\varepsilon _{-}\left( q \right)}\frac{1}{q-\left( {{k}_{x}}+i\epsilon  \right)}.
	\end{gather}
	
	The behavior of electric field in the immediate vicinity of the junction, $x \rightarrow +0$, can be understood by analyzing the decay of Fourier components at large $q$. The spectrum of metal-contacted field (\ref{eq-er-metallized}) rapidly decays as $q^{-3/2}$, which implies that the field at the junction is zero. This is quite expected, as the electric field in metal is absent, while its tangential component is continuous across the interface. The electric field spectrum for terminated junction (\ref{eq-er-terminated}) decays $q^{-1}$. It implies that the junction field is finite. Its value can be linked to the residue at infinity
	\begin{equation}
		E^{\times}\left( x=0 \right)=\underset{q\to \infty }{\mathop{\lim }}\,\left[ iqE_{R}^{\times}\left( q \right) \right]=\frac{{{E}_{0}}{{\varepsilon }_{-}}\left( {{k}_{x}} \right)}{1+{{\eta }}/\sin \theta }.
	\end{equation}
	A fully analytical result can be obtained for normal incidence, $k_x=0$. In that case, the principal value part of Cauchy integral (\ref{eq-cauchy-factorization}) is zero by the virtue of integrand anti-symmetry. The pole part of the Cauchy integral is evaluated trivially, which results in $\varepsilon_{-} \left( q = 0 \right)= \varepsilon^{1/2}_{-} \left( q = 0 \right)$. Noteworthy is the resulting 'anomalous' screening of the incident field:
	\begin{equation}
		E^{\times}\left( x=0 , k_x = 0\right)=\frac{E_0}{\sqrt{1+{{\eta }}} }.
	\end{equation}
	It contrasts to the linear screening of the incident field by an extended 2D electron system, $E = E_0/(1+\eta)$. Queerly speaking, truncating a half of 2D system results in halving of the power in the screening law.
	
	\begin{figure}[ht!]
		\includegraphics[width=0.9\linewidth]{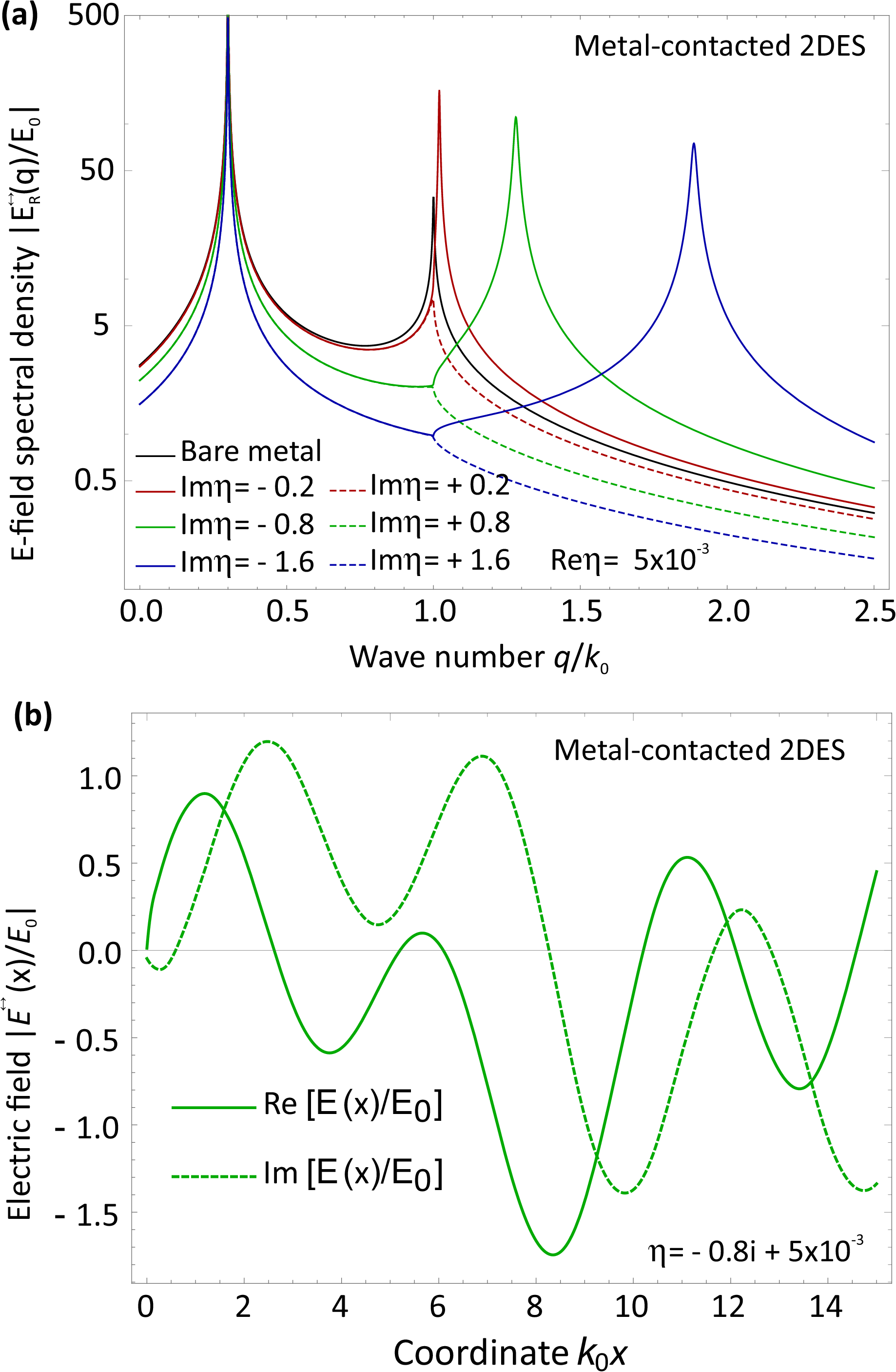}
		\caption{(a) Spectral structure of the diffracted fields, i.e. the dependences of diffracted field $E_R(q)$ on the wave number at different values of surface conductivity $\eta$. Solid and dashed lines correspond to the capacitive and inductive conductivities, respectively (b) Real-space structure of the diffracted field in case of capacitive 2d conductivity $\eta'' = -0.8$. Both plots are presented for metal-contacted 2DES with $\eta'=5\times 10^{-3}$, $k_x = 0.3 k_0$}
		\label{fig:spectra_and_real_space}
	\end{figure}
	
	It is instructive to reveal the singularities of the field spectra $E_R(q)$. They are linked to the waves emitted upon the diffraction. Figure \ref{fig:spectra_and_real_space} (a) shows the spectra of electric field $E_R(q)$ for metal-contacted 2DES with different values of surface conductivity. The external wave is incident with $k_x = 0.3 k_0$. Naturally, all spectral curves have a pole at $q = k_x$ that corresponds to the incident field screened by the 2DES. Another (weaker) singularity is present at $q = k_0$ in the absence of 2DES ($\eta=0$, black line). The spectrum of electric field behaves as $E_R(q) \sim (k_0-q)^{1/2}$. This corresponds to the radiation of a 'line of dipoles' arranged along the contact line. In real space, such a weak singularity corresponds to the cylindrical wave $E(x) \sim |x|^{-1/2}e^{ik_0 x}$. With enhancement in 2DES conductivity, this 'dipole radiation' is largely modified. For inductive 2d conductivity, $\eta''>0$, the singularity is smeared, and all its traces fade at $\eta'' \sim 1$. We can speculate that inductance of 2d electrons prevents the propagation of dipole radiation.
	
	For capacitive 2d conductivity, $\eta''<0$, the spectral features are more interesting. An initial weak singularity at $q=k_0$ shifts rightwards and becomes a well-developed pole. The width of spectral peak remains finite only by the virtue of dissipative conductivity, $\eta' \neq 0$. The situation corresponds to the launching of transverse electric two-dimensional plasmons by the edge. Figure \ref{fig:spectra_and_real_space} (b) illustrates this situation in real space: a slow modulation of electric field with period $k_x^{-1}$ is threaded with faster modulation due to the plasma wave. The plasmon wavelength shrinks with increasing $|\eta''|$.
	
	A more detailed study of the real-space electric field can be achieved by inverse Fourier transforming Eq.~(\ref{eq-solution}):
	\begin{equation}
		{{E}_{R}}\left( x \right)=\frac{1}{2\pi }\int\limits_{-\infty }^{+\infty }{dq{{E}_{R}}\left( q \right){{e}^{iqx}}}.
	\end{equation}
	\rev{To ensure the decay of Fourier exponent $e^{iqx}$ at $x>0$, we close the integration contour for $E_R(x)$ in the {\it upper} half plane, blue line in Fig.~\ref{fig:structure} (b). We note here that $E_R(q)$ is non-analytic in the UHP, still, its singularities are readily recognized. The recognition is achieved by writing $\varepsilon_{-}(q) = \varepsilon(q)/\varepsilon_{+}(q)$ in the denominators of Eqs.~(\ref{eq-er-metallized}) and (\ref{eq-er-terminated}). With that substitution, all singularities of $E_R(q)$ in the UHP reduce to (1) a pole at $q=k_x+i\epsilon$ corresponding to the incident wave screened by the 2DES (2) a pole at the zero of 2DES dielectric function $\varepsilon(q)$ (3) a branch cut of $\varepsilon(q)$ running from $+k_0$ to $+i\infty$. It is possible to show that the TE mode pole \# 2 exists onlt at $\eta''<0$ and is located at}
	\begin{equation}
		\label{eq-te-pole}
		q_{\rm pl} = k_0 \sqrt{1 - \eta^2}.
	\end{equation}
	\rev{The seeming independence of $q_{\rm pl}$ on the sign of $\eta''$ is spurious and appears due to developing a square of two terms comprising $\varepsilon(q)$. From an original definition of $\varepsilon(q)$, Eq.~\ref{eq-dielectric-function}, it is clear that its zeros can exist if only $\eta''<0$. The three singularities of $E_R(q)$ in the UHP produce the three contributions to the real-space field in the 2DES, respectively}
	\begin{equation}
		{{E}_{R}}\left( x \right)	=\frac{E_0 {e}^{i{{k}_{x}}x}}{1 + \eta/\sin\theta} + i{{e}^{i{{q}_{\rm pl}}x}}\underset{q={{q}_{\rm pl}}}{\mathop{\text{Res}}}\,{{E}_{R}}\left( q \right)-{E_{\rm b.c.}(x)}.
	\end{equation}
	\rev{The last term, $E_{\rm b.c.}(x)$ is the result of inverse Fourier transforming the field spectrum along the branch cut of the dielectric function $\varepsilon_R(q)$. The resulting contribution to the electric field is rapidly evanescent already at $x\sim k_0^{-1}$. The TE mode makes a longer run, least for weakly dissipative conductivity. As the long-range field is largely determined by this wave, it is natural to study its amplitude $E_{\rm pl}$ in more detail.}
	
	\begin{figure}[ht!]
		\centering
		\includegraphics[width=0.85\linewidth]{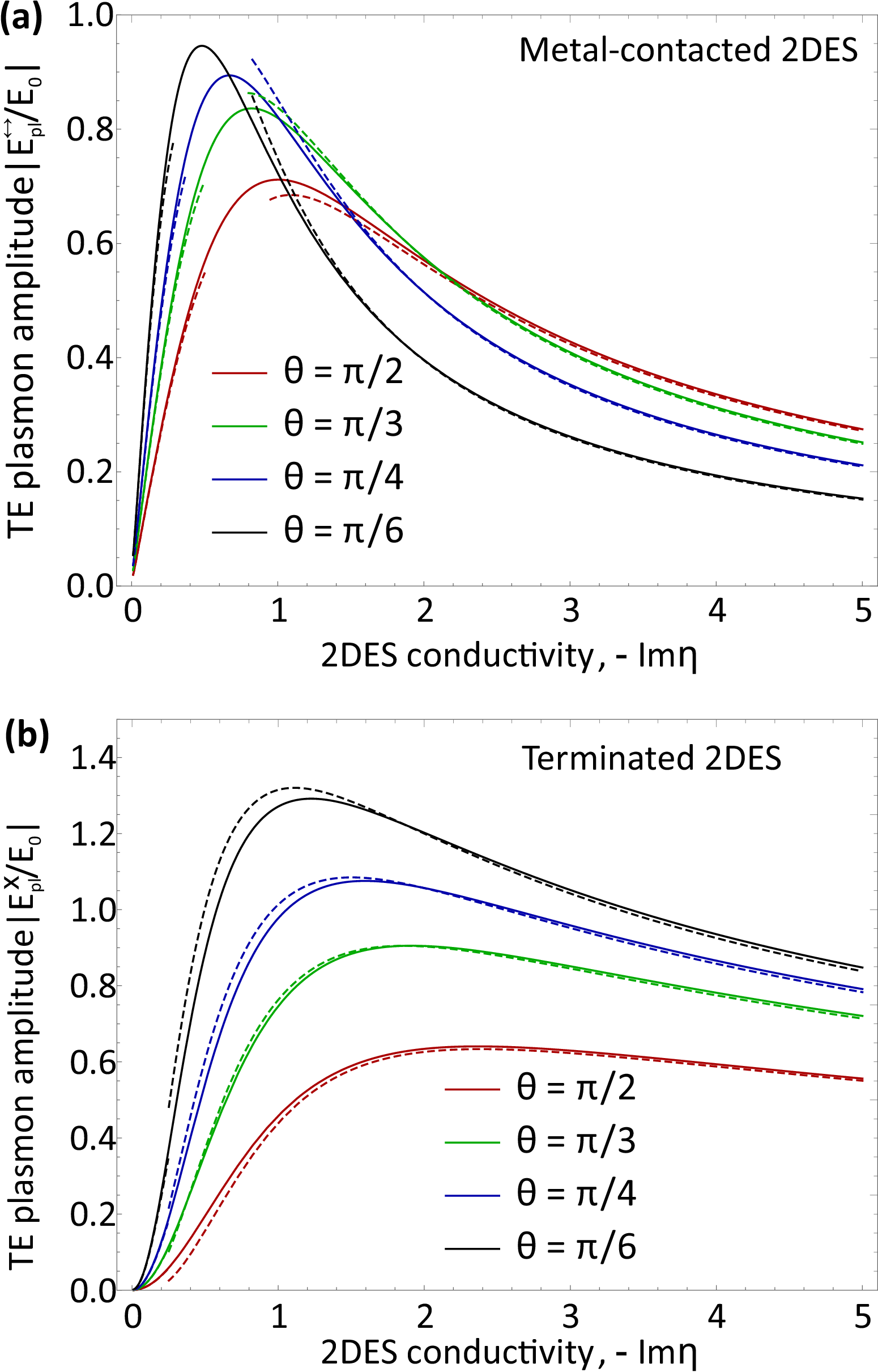}
		\caption{Conversion efficiency of free-space radiation into 2d plasmons for metal-contacted (a) and terminated (b) 2DES vs 2DES conductivity $-\eta''$ at various incidence angles $\theta$. Solid lines are obtained with exact expressions (\ref{eq-plasm-terminated}) and (\ref{eq-plasm-metallized}), dashed lines are obtained with simplified asymptotic expressions (\ref{eq-plasm-low-sigma}) for low conductivity and (\ref{eq-plasm-high-eta}) for high conductivity. }
		\label{fig:plasmons}
	\end{figure}
	
	An explicit evaluation of the residue at $q=q_{\rm pl}$ results in the following expressions for TE mode amplitude for terminated 2DES
	\begin{equation}
		\label{eq-plasm-terminated}
		E_{pl}^{\times}=\frac{{{E}_{0}}}{1+{\eta}/\sin \theta }\frac{\varepsilon _{-} \left( {{k}_{x}} \right)\varepsilon _{+}\left( {q_{pl}} \right)}{{{\left. \partial \varepsilon\left( q \right)/\partial q \right|}_{q={q_{pl}}}}}\frac{1}{{q_{pl}}-{{k}_{x}}},
	\end{equation}
	and metal-contacted 2DES
	\begin{equation}
		\label{eq-plasm-metallized}
		E_{pl}^{\leftrightarrow}=E_{pl}^{\times}\frac{\sqrt{{{k}_{0}} -{{k}_{x}}}}{\sqrt{{{k}_{0}} -{{q}_{pl}}}}.
	\end{equation}
	The conversion coefficients of free-space radiation to TE modes (\ref{eq-plasm-terminated},\ref{eq-plasm-metallized}) are shown in Fig.~\ref{fig:plasmons} as a function of $-\eta''$ at various angles of incidence $\theta$. The dissipative part of conductivity is hereby assumed very small,  $\eta' = 5\times 10^{-3}$. The conversion coefficients approach zero for small conductivity linearly (for metal-contacted 2DES) and quadratically (for terminated 2DES). This scaling can be shown analytically by recalling that all dielectric functions approach unity as $\eta \rightarrow 0$. In this limit, we find
	\begin{gather}
		\label{eq-plasm-low-sigma}
		E_{pl}^{\times}=\frac{{E_0}}{2}\frac{{{\left| {{\eta }''} \right|}^{2}}}{{{\sin }^2}\frac{\theta }{2}},\\
		E_{pl}^{\leftrightarrow}={E_0}\left| \frac{{\eta''}}{\sin \frac{\theta }{2}} \right|.
	\end{gather}
	These asymptotes, shown in Fig.~\ref{fig:plasmons} with dashed lines near the origin, agree well with full computed conversion coefficients in the limit $-\eta'' \ll 1$.
	
	We finally provide the conversion coefficients in the opposite limit $|\eta''| \gg 1$. The complexity lies in evaluation of factorized functions $\varepsilon_{\pm}$ at $q=k_x$ and $q=q_{pl}$. These functions are now largely different from unity, while the integrand in Cauchy factorization (\ref{eq-cauchy-factorization}) is full of singularities. To cope with the problem, we upraise all zeros and singularities from the definition of dielectric function:
	\begin{gather}
		\varepsilon(q)= \frac{q^2 - q_{pl}^2}{\eta k_0 \sqrt{k_0^2 - q^2} f_{\rm aux}(q)},\\
		f_{\rm aux}(q) = 1 + \frac{1}{\eta}\frac{\sqrt{k_0^2 - q^2}}{k_0}.
	\end{gather}
	The Cauchy factorization of a zero-free non-singular function $f_{\rm aux}(q)$ is much simpler. Moreover, if we are interested only in the absolute value of TE mode amplitude, ignoring the phase, only a selected section of Cauchy integral should be considered, namely
	\begin{equation}
		|f_{\rm aux\pm}(q)| = \sqrt{|f_{\rm aux}(q)|} e^{\pm I},
	\end{equation}
	\begin{multline}
		\label{eq-plasm-high-eta}
		I = \frac{1}{2\pi}\int\limits_{-1}^{+1}\frac{{\rm arctan}\left(\frac{\sqrt{1-u^2}}{|\eta''|}\right)}{u-q/k_0}du\approx\\
		-\frac{1}{2\left| {\eta ''} \right|{{k}_{0}}}
		\left\{ \begin{aligned}
			& q,\,\,q<{{k}_{0}} \\ 
			& q-\sqrt{{{q}^{2}}-1},\,\,q>{{k}_{0}} \\ 
		\end{aligned} \right\}.	
	\end{multline}
	
	Asymptotic values of TE mode amplitude obtained with above approximate technique agree very well with those computed exactly. Surprisingly, the approximate scheme works fine even for $|\eta ''| \lesssim 1$, as shown in Fig.~\ref{fig:plasmons} with dashed lines.
	
	In conclusion, we compare the peculiarities of TE-wave diffraction with those of TM-diffraction studied previously~\cite{Nikulin_Edge}. Electric field for TM diffraction is enhanced at $x=0$ in a singular fashion, $E_{TM}(x=0) \approx E_0 [\eta(1+\eta)]^{-1/2}$. Such enhancement is an aftersound of the lightning rod effect at the keen non-contacted metal edge. The lightning rod effect is polarization-selective. For electric field along an edge, it turns into field suppression by dynamic currents in metal. Strong and highly non-uniform electric field for TM polarization resulted in very large photon-to-plasmon conversion efficiency which scaled as $|\eta''|^{-1/2}$ for low surface conductivity. Relative smoothness of diffracted field for TE-polarization results in moderate amplitudes of the launched plasma waves. At $|\eta''|\sim 1$, ultimate values of plasmon amplitude $E_{pl} \sim E_0$ are reached both for terminated and metal-contacted 2DES. While all electric field components in the considered polarizations remain finite, the magnetic field (being proportional to electric gradients) diverges in the vicinity of terminated edge. This can result in high efficiency of 2d magnon launching~\cite{2d_magnons,Nikitov_Magnons}, as well as to the local enhancement of electron spin resonance signals~\cite{2d_ESR}.
	
	The presented analytical method is readily generalized to the 2DES located above perfect conductors (gates). This is achieved via replacements $\eta_{L/R} \rightarrow \eta_{L/R}(1-e^{-2d\sqrt{q^2-k_0^2}})$, where $d$ is the distance to the gate. With such a modification, it will be possible to study the hybridization of TE plasmons with cavity modes formed between 2DES and its gate~\cite{Muravev_grating}. Previous studies of scattering problems in such spatially non-uniform gated 2DES were limited to the weak non-locality approxiations~\cite{Rodionov_WeakNL,zagorodnev_confined}.
	
	The current study dealt with an electromagnetic scattering problem. Another class of EM problems deals with the properties of edge eigenmodes~\cite{volkov1988edge,Sokolik_inclined,Margetis_analytical}, which remain largely unexplored for capacitive materials with $\eta''<0$. The edge TE plasmons, if exist, would have two non-trivial components of electric field. This would complicate the Wiener-Hopf analysis of the spectra~\cite{zabolotnykh2016edge}. Fortunately, efficient factorization methods for such matrix-type Wiener-Hops systems have recently emerged~\cite{daniele2014wiener}.
	
	All our consideration relied on the local model of 2DES conductivity. This is justified by relative smoothness of emerging fields, least in the limit $|\eta ''|\lesssim 1$. An interesting aspect of conduction non-locality in TE polarization is the possible launching of shear waves~\cite{Afanasiev_Shear,Falkovich_Shear}. Such waves can propagate due to electron viscosity normally to the junction; the non-uniformity of the diffracted field provides the necessary momentum. Therefore, metal-2DES junctions illuminated with TE-polarized waves may represent a platform for studies of exotic shear excitations.
	
	The obtained solution for local fields generated upon diffraction at the junction can be used as a building block for modeling of 2d photodetectors. The underlying light-to-current conversion mechanism can be photo-thermoelectric effect~\cite{Levitov_Hot_Theory}, photovoltaic effect~\cite{Echtermeyer_MGJ}, or photon drag~\cite{Durnev_structured,Entin_PD}. In all these cases, the photocurrent is a known quadratic functional of the local eclectic fields, whose particular values can be computed with our main Eqs.~(\ref{eq-solution}). Such scheme of photocurrent computation is simpler and more transparent than EM simulations used previously~\cite{Olbrich_Ratchet,Fateev_Rectification}. It can also provide analytical insights into the limiting performance of 2d optoelectronic devices.
	
	The work was supported by the grant \# 21-79-20225 of the Russian Science Foundation.
	
	\bibliography{sample}

\end{document}